%% file: confineknots-v2.2-arxiv.tex
\documentclass[aps,preprint,showpacs,superscriptaddress,groupedaddress]{revtex4}  

\usepackage{etex}
\usepackage{graphicx}  
\usepackage{dcolumn}   

\usepackage{bm}        
\usepackage{amssymb}   
\usepackage{amsmath}
\usepackage{threeparttable}

\usepackage{caption}
\usepackage{verbatim}
\usepackage{rotating}

\def\fiverm{\footnotesize}
\input prepictex.tex
\input pictex.tex
\input postpictex.tex

\usepackage[usenames,dvipsnames,svgnames]{xcolor}

\input macros

\begin{document}

\widetext
\leftline{Version 2.1 as of \today}

\title[Relative knot probabilities in confined lattice polygons]{
Relative knot probabilities in confined lattice polygons}

\author{EJ Janse van Rensburg}
\affiliation{Department of Mathematics and Statistics, 
York University, Toronto, Ontario M3J~1P3, Canada (Email: rensburg@yorku.ca)}
\author{E Orlandini}
\affiliation{Dipartimento di Fisica e Astronomia e Sezione INFN, Universit\`a di Padova, 
Via Marzolo 8, I-35131 Padova, Italy (Email: orlandini@pd.infn.it)}
\author{MC Tesi}
\affiliation{Dipartimento di Matematica, Universit\`a di Bologna, \\ Piazza di Porta San Donato 5, I-40126 Bologna, Italy (Email: mariacarla.tesi@unibo.it)}

\date{\today}

\begin{abstract}
In this paper we examine the relative knotting probabilities in a lattice model 
of ring polymers confined in a cavity.  The model is of a lattice knot of size $n$
in the cubic lattice, confined to a cube of side-length $L$ and with volume
$V=(L{+}1)^3$ sites.  We use Monte Carlo algorithms to approximately enumerate
the number of conformations of lattice knots in the confining cube.  
If $p_{n,L}(K)$ is the number of conformations of a lattice polygon of length $n$
and knot type $K$ in a cube of volume $L^3$, then the relative knotting
probability of a lattice polygon to have knot type $K$, relative to the 
probability that the polygon is the unknot (the trivial knot, denoted by $0_1$), is 
$\rho_{n,L}(K/0_1) = p_{n,L}(K)/p_{n,L}(0_1)$.  We determine $\rho_{n,L}(K/0_1)$
for various knot types $K$ up to six crossing knots.  Our data show that
these relative knotting probabilities are small so that the model is dominated
by lattice polygons of knot type the unknot.  Moreover, if the concentration
of the monomers of the lattice knot is $\varphi = n/V$, then the relative
knot probability increases with $\varphi$ along a curve that flattens off as
the Hamiltonian state is approached. 
\end{abstract}

\maketitle

\onecolumngrid

\section{Introduction}

The physical properties of polymers and biopolymers are affected by knotting
and linking  between polymeric strands \cite{D62,deG84,KM91,BO12}.  
In reference \cite{GFR96} a Flory
theory for knotting ring polymers was proposed, and the probability of knotting
in DNA were examined in references \cite{RCV93} and \cite{SW93}.   The effects
of confinement on the knotting and entanglements of polymeric strands have
been modelled using polygonal models \cite{DEMRZ14,ERZ21} and sampling
randomly generated polygons in confining spaces (see, for example, the
algorithms in references \cite{DEMZ11,DEMZ12,DEMZ12a,DERZ18,CSS24}).  
Knot probabilities have also been examined in randomly generated plane diagrams
\cite{CCM16,W20}, and other flat polygonal structures \cite{OSV04}.
In these studies the consensus is that confining a polymer will increase
the probability that it is knotted when undergoing ring closure \cite{CD86,AVTSR02}. 
In the case of extremely strong confinement quantum-inspired coding was 
designed and used, for the first time, to characterise the entanglement 
of topologically unrestricted ring polymers at maximum packing density \cite{SHFM23}.
More results on the knotting probability in models of ring polymers are contained
in references \cite{MW84,KM91,VLKFC92,JvR02,JvRR11A}. 

There is also a comprehensive literature on the use of lattice models to 
examine the entanglement complexity of polymers.  These are self-avoiding walk and 
lattice polygon models, and they model the self-avoidance of polymeric strands 
while also representing the topological complexity of polymers 
\cite{MW84,JvRW91,KM91,JvR02} (also in the context of being squeezed
into small volumes \cite{JvR07,MLY11}).  A lattice polygon being stretched in
one direction by a force, or adsorbed onto a hard wall, has reduced probability 
of knotting \cite{OSV04,JOTW07,JOTW07b,V95}, but when it has transitioned 
through the $\theta$-point into a collapsed phase its knotting probability is 
increased \cite{TJvROSW94}.  It is also known that thermodynamic quantities
are functions of knotting \cite{JvR07,GJvRR12,JvR14}, and the osmotic
pressure of compressed lattice knots is a function of knot type for 
finite size polygon models of ring polymers \cite{JvR19}. 

\begin{figure}[h!]
\includegraphics[width=0.45\textwidth]{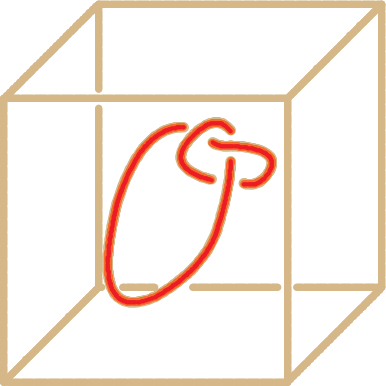}
\captionsetup{justification=raggedright,singlelinecheck=false}
\caption{A schematic drawing of a knot in a confining cube or box.  The
side length of the cube is $L$, and the knot has rotational, translational and
conformational degrees of freedom.  These contribute to the free energy
of the system. Notice that the cube contains $(L{+}1)^3$ lattice sites
so that its volume is $V=(L{+}1)^d$ sites.}
\label{1}  
\end{figure}

In this paper we use a lattice knot model of a ring polymer in a confined space to
model the dependence of its knotting probability on its knot type.  The
lattice knot is compressed inside a confining cube, and also is a model of a
biopolymer in a cavity or in a space between membranes.
The compression and entanglement of single DNA molecules have been 
experimentally investigated \cite{TDD11,R15}, and a lattice model of knotting
in such confining conditions may give some insights into the results of these
experiments even as the lattice may be an inadequate approximation.  The aim here
is to determine some quantitative results in the lattice to get qualitative 
insights into a complex physical situation, in particular insofar the relative
incidence of knotting as a measure of entanglement of the polymeric molecules.

In Figure \ref{1} a schematic diagram of the model is shown.  The free energy
of the model is quantified by the rotational, translational and conformational 
degrees of freedom of the knot inside the cube.  In the cubic lattice the knot
is a (self-avoiding) lattice polygon of knot type $K$, and its entropy is quantified 
by introducing the function $p_{n,L}(K)$ which is the number of conformations
of the lattice knot if has length $n$ and knot type $K$, and the confining 
cube has side length $L$ and has $V=(L{+}1)^3$ lattice sites.

In section 2 we give a brief review of knot entropy and related issues.  We
will in particular be interested in the number of conformations of a lattice polygon
of fixed knot type $K$.  In section 3 we discuss the sampling of lattice knots
using the GAS and GARM algorithms \cite{JvRR09,JvRR11A,RJvR08}, and in section 4 
we present and discuss the results of simulations.  The paper is concluded in 
section 5 with a few final remarks.

\begin{figure}[h!]
\includegraphics[width=0.45\textwidth]{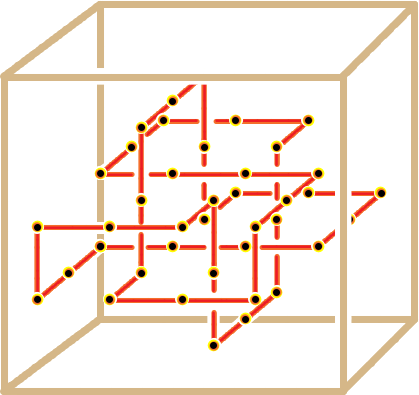}
\captionsetup{justification=raggedright,singlelinecheck=false}
\caption{A lattice knot in a cube in the cubic lattice.  The side length of the cube is 
$L-1$ (so that it contains $V=L^3$ sites).}
\label{2}  
\end{figure}

\section{Lattice knots}

The number of cubic lattice \textit{polygons} of length $n$ is denoted 
$p_n$ (these are unrooted and closed self-avoiding walks of length 
$n$ steps in the cubic lattice counted up to equivalence under translations 
in the lattice).  A \textit{lattice knot} is a lattice polygon of a given (fixed) 
knot type $K$, and the number of lattice knots of length $n$ and knot type 
$K$ is denoted by $p_n(K)$.

A cubic lattice knot in a confining cube is shown in figure \ref{2}.  The
side-length of the cube is $L$ in terms of lattice steps so that the
cube contains $(L{+}1)^3$ lattice sites.  The knot type of a lattice knot
realised inside the cube (so that it is a cycle in the cube) is fixed,
and so this is a model of a knotted ring polymer inside a confining volume,
with fixed knot type.  The lattice knot also has fixed length $n$ (edges),
and its distinct conformations (also due to translations and rotations
inside the cube) is a model of the configurational entropy of a 
confined ring polymer.  This model was explored in 
references \cite{JvR19,JvR20}, where its free energy was
modelled by Flory-Huggins theory, and the  Flory Interaction Parameter
of the model was estimated to be $\chi = 0.18(3)$ \cite{JvR20}.

If $K=0_1$ (the unknot), then $p_4(0_1)=3$ in the cubic lattice,
and $p_n(0_1)=0$ if $n<4$.  This shows that the \textit{minimal length}
$n_{0_1}$ of the unknot in the cubic lattice is $4$ \cite{deG84}.  
The minimal length of lattice knots of type $K$ \cite{JvRP95,JvRR11} is 
denoted $n_K$, and for the trefoil knot $3_1$, $n_{3_1}=24$ \cite{D93}.
For the figure eight knot $4_1$, $n_{4_1}=30$ and for the knot $5_1$,
$n_{5_1}=34$ \cite{SIAD09}. Numerical estimates of the minimal 
length of $n_K$ of other knot types have also been determined, and for
knots of low crossing number, those are exact with high probability
(although unproven).

Since the cubic lattice is bipartite all lattice polygons have even lengths, so that 
$p_n(K)=0$ if $n$ is odd.  By concatenating two lattice unknots (see 
reference \cite{SW88}), it follows that, by Fekete's lemma, the limits
\begin{equation}
\lim_{n\to\infty} (p_n(0_1))^{1/n} = \mu_{0_1} < \mu
= \lim_{n\to\infty} p_n^{1/n}
\end{equation}
exist if these are taken through even values of $n$, where $\mu$ is 
the \textit{growth constant} of self-avoiding cubic lattice polygons (or of 
the cubic lattice self-avoiding walk), and $\mu_{0_1}$ is the growth constant 
of lattice unknots.  That $\mu_{0_1}<\mu$ was shown in reference \cite{SW88}.

Generally, the growth constant of cubic lattice knots of type $K$ is defined by
\begin{equation}
\limsup_{n\to\infty} (p_n(K))^{1/n} = \mu_K .
\end{equation}
It is known that $\mu_{0_1}\leq \mu_K < \mu$ (see, for example,
\cite{SW88,JvR99}), and it is believed that $\mu_K=\mu_{0_1}$ for 
all knot types $K$ \cite{JvR02,JvR09}.  There are strong numerical evidences
that
\begin{eqnarray}
p_n &=& C\, n^{\alpha-3}\, \mu^n\,(1+o(1))
\label{eqn3a} \\
p_n(K) &=& C_K\, n^{\alpha+N_K-3}\, \mu_K^n\,(1+o(1))
\label{eqn3}
\end{eqnarray}
where $C_K$ is an amplitude dependent on the lattice and the knot type
$K$, $\alpha=0.237(2)$ \cite{GZJ98,LGZ89} is the entropic exponent of lattice 
polygons and $N_K$ is the number of prime knot components in the 
knot $K$ \cite{OTJW96}.

Lattice knots in $\IntZ^3$ were sampled using both the GARM \cite{RJvR08}
and GAS \cite{JvRR09,JvRR11A} algorithms, implemented with BFACF 
elementary moves \cite{BF81,ACF83,JvRR08,JvRR12}.  Both these
algorithms are approximate enumeration algorithms, returning estimates 
of $p_n(K)/p_m(K)$.  Choosing $m$ equal to the minimal length of the lattice 
knot of type $K$, and exactly counting $p_m(K)$, estimates of $p_n(K)$ for 
$n\geq m$ are obtained. Normally, one chooses $m=n_K$ and
uses exact counts for $p_m(K)$ to determine estimates of $p_n(K)$ for
$n>n_K$.  For example, by equation
\Ref{eqn3} one expects $n\,p_n(0_1) \sim p_n(3_1)$.  Generally, one expects
that there is a value of $\ell$ such that
\begin{equation}
p_{n-\ell} (0_1) \simeq n^{-1}\,p_n(K), \quad\hbox{as $n\to\infty$},
\end{equation}
for a given prime knot type $K$.  Plotting 
$\log \left[ n\,p_{n-\ell}(0_1)/ p_n(K) \right]$ as a function of $1/n$ for various
values of $\ell$ shows that $\ell\approx 8$ when $K=3_1$ is the trefoil knot
\cite{JvRR11A,BO12}.  This shows that tying a trefoil knot in a lattice polygon 
reduces the entropy of the polygon by reducing its length by about $8$ steps.  
In addition, one notices that $p_n(0_1) > p_n(3_1)$ for small $n$, but
these results show that $p_n(0_1) \approx p_n(3_1)$ if $n\approx  169600\pm5600$
and that $p_n(0_1)<p_n(3_1)$ for $n$ larger than this, consistent with
equation \Ref{eqn3} \cite{JvRR11A}.

The probability of knotting in polygons of length $n$ sampled uniformly is
\begin{equation*}
{\mathbb P}_n = 1 - p_n(0_1)/p_n = 1 - (C_{0_1}/C)\,(\mu_{0_1}/\mu)^n
(1+o(1)) ,
\end{equation*}
by equations \Ref{eqn3a} and \Ref{eqn3}.  Since $\mu_{0_1}<\mu$ this 
shows that ${\mathbb P}_n \to 1$ as $n\to\infty$ (this is the Sumners-Whittington
result in reference \cite{SW88}).  Numerical simulations in reference 
\cite{JvRW90} shows that $\log (\mu/\mu_{0_1}) = (4.15\pm 0.32) \times 10^{-6}$.

\subsection{Lattice knots confined to cubes in the cubic lattice}

Confined lattice knots have reduced state spaces dependent on the size 
of the confining cube. Decreasing the size of the confining cube reduces 
the conformational and, for short lattice knots, also the rotational and 
translational, degrees of freedom of the lattice knot.  This models the
reduction in entropy when a knotted ring polymer is compressed in a cavity. 

Denote the number of placements of lattice polygons (as opposed 
to lattice knots) of length $n$ in a cube of side-length $L$ 
(this is an \textit{$L$-cube}) by $p_{n,L}$.  
The number of lattice knots of length $n$ and knot type $K$, confined to 
an $L$-cube of sidelength $L$ containing $(L{+}1)^3$ lattice sites (as 
illustrated in figure \ref{2}), is denoted by $p_{n,L}(K)$.  For example,
for the unknot with $n=4$ and $L=1$, $p_{4,1}(0_1)=6$  (these are minimal 
length unknots of length $b$ in a cube of sidelength $L=1$.
Notice that, in general, both conformational and 
translational degrees of freedom contribute to the counts $p_{n,L}(K)$.  There 
are few exact results available for $p_{n,L}(K)$, but clearly,
\begin{equation}
p_{n,L}(K) = 0,\qquad\hbox{if $n<n_K$ or $n>L^3$.}
\end{equation}
For minimal length lattice knots and various knot types, $p_{n,L}(K)$
can be evaluated by exhaustively counting 
their number of placements in $L$-cubes.  These are shown
in table \ref{table1} (these counts are, strictly speaking, lower bounds, but should
be exact in most cases, as they were checked exhaustively by 
repeated computer searches).

\begin{table}[h]
\begin{threeparttable}
\caption{Table I: Number of lattice knots $p_{n,L}(K)$ of 
minimal length $n_K$ in an $L$-box}
\setlength{\extrarowheight}{1pt}
\setlength{\tabcolsep}{2.5pt}            
\begin{tabular}{lrrrrrrrrrr}
\hline
$L$-Box & $0_1$ & $3_1^+$ & $4_1$ & $5_1^+$ & $5_2^+$ & $6_1^+$ & $6_2^+$ & $6_3$ 
& $3_1^+\#3_1^+$ & $3_1^+\#3_1^-$ \\
\hline
$1$ & $6$ & $0$ & $0$ & $0$ & $0$ & $0$ & $0$ & $0$ & $0$ & $0$ \\ 
$2$ & $36$ & $0$ & $0$ & $0$ & $0$ & $0$ & $0$ & $0$ & $0$ & $0$ \\
$3$ & $108$ & $2084$ & $864$ & $0$ & $0$ & $0$ & $0$ & $0$ & $0$ & $108$ \\
$4$ & $240$ & $15052$ & $14048$ & $9840$ & $123312$ & $6144$ & $14568$ & $4032$ & $1332$ & $33192$ \\
$5$ & $450$ & $48876$ & $73440$ & $52416$ & $740232$ & $37512$ & $134928$ & $32496$ & $61452$ & $634752$ \\
$6$ & $756$ & $113540$ & $188928$ & $147504$ & $2186976$ & $112329$ & $459480$ & $106272$ & $222456$ & $2201904$ \\
$7$ & $1176$ & $219028$ & $386400$ & $315120$ & $4808280$ & $249000$ & $1086720$ & $246672$ & $530280$ & $5166360$ \\
$8$ & $1728$ & $375324$ & $687744$ & $575280$ & $8948880$ & $465984$ & $2115144$ & $475008$ & $1030860$ & $9959832$ \\
$9$ & $2430$ & $592412$ & $1114848$ & $948000$ & $14953512$ & $781704$ & $3643248$ & $812592$ & $1770132$ & $17014032$ \\
$10$ & $3300$ & $880276$ & $1689600$ & $1453296$ & $23166912$ & $1214592$ & $5769528$ & $1280736$ & $2794032$ & $26760672$ \\
$11$ & $4356$ & $1248900$ & $2433888$ & $2111184$ & $33933816$ & $1783080$ & $8592480$ & $1900752$ & $4148496$ & $39631464$ \\
$12$ & $5616$ & $1708268$ & $3369600$ & $2941680$ & $47598960$ & $2505600$ & $12210600$ & $2693952$ & $5879460$ & $56058120$ \\
$13$ & $7098$ & $2268364$ & $4518624$ & $3964800$ & $64507080$ & $3400584$ & $16722384$ & $3681648$ & $8032868$ & $76472352$ \\
$14$ & $8820$ & $2939172$ & $5902848$ & $5200560$ & $85002912$ & $4486464$ & $22226328$ & $4885152$ & $10654632$ & $101305872$ \\
$15$ & $10800$ & $3730676$ & $7544160$ & $6668976$ & $109431192$ & $5781672$ & $28820928$ & $6325776$ & $13790712$ & $130990392$ \\
\hline
$n_K$ & $4$ & $24$ & $30$ & $34$ & $36$ & $40$ & $40$ & $40$ & $40$ & $40$ \\
\hline
\end{tabular}
\label{table1}
\end{threeparttable}
\end{table}

\begin{figure}[h!]
\includegraphics[width=0.5\textwidth,height=0.33\textheight]{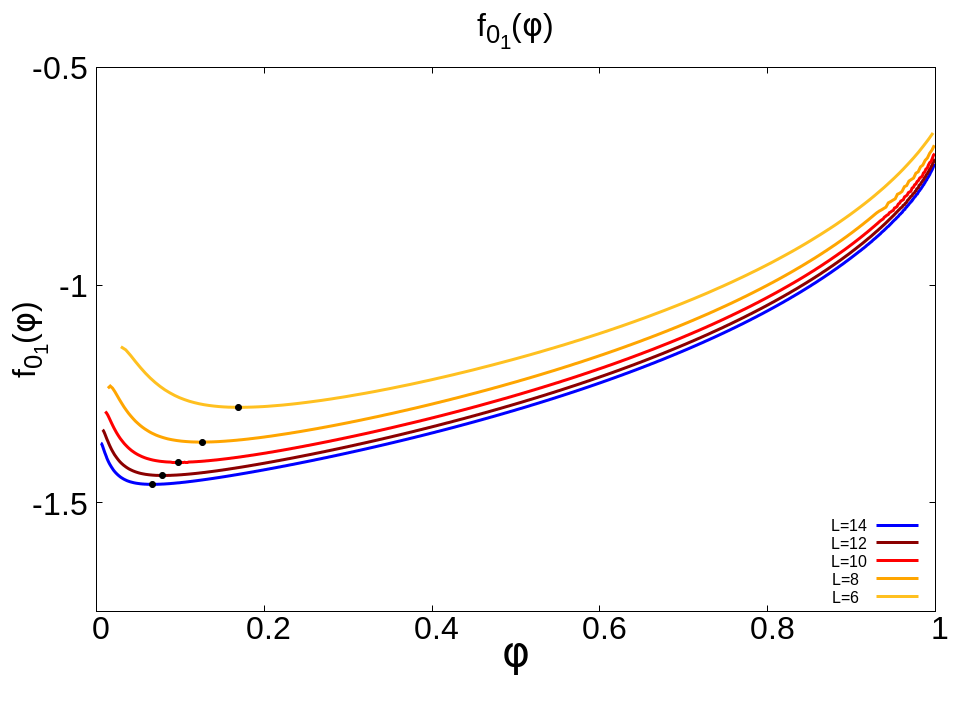}
\captionsetup{justification=raggedright,singlelinecheck=false}
\caption{Free energy curves of lattice unknots (see equation \Ref{88})
for $L\in\{6,8,10,12,14\}$.  The curves appear to converge to a 
limiting curve with increasing $L$ and have minima at a critical
concentration $\varphi_c$ (dependent on $L$).  The minima are marked 
by a bullet on each curve.}
\label{figf01}
\end{figure}

\begin{figure}[h!]
\includegraphics[width=0.4\textwidth,height=0.225\textheight]{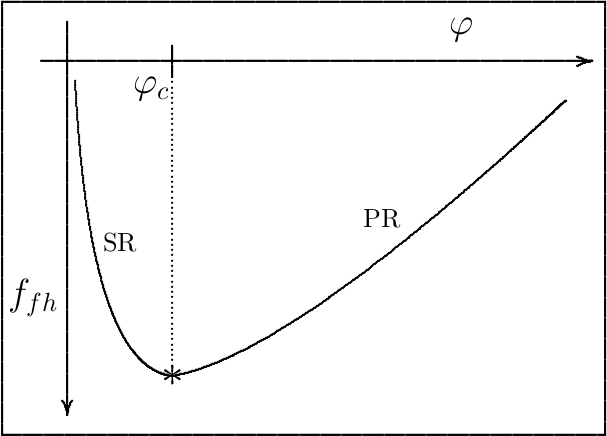}
\captionsetup{justification=raggedright,singlelinecheck=false}
%
%
%
%
%
%
\captionsetup{justification=raggedright,singlelinecheck=false}
\caption{A schematic representation of the Flory-Huggins free energy.
The convex curve has a minimum at $\varphi_c$, separating the model
into two regimes, namely a solvent-rich (SR) phase at low concentrations
$\varphi<\varphi_c$, and a polymer-rich (PR) phase at high concentrations
$\varphi>\varphi_c$.  The regimes are separated at the critical concentration
$\varphi_c$.  In the vicinity of $\varphi_c$ the system is in $\theta$-conditions.}
\label{G1}
\end{figure}

The free energy of lattice knots of type $K$ in an $L$-cube was examined 
numerically in references \cite{JvR19,JvR20}.  The lattice free energy 
density of compressed lattice knots can be defined by 
\begin{equation}
F_V(\varphi) = - \sfrac{1}{V} \log p_{n,L}(K)
\end{equation}
where $V=(L{+}1)^3$ is the number of sites of the confining $L$-cube, and 
$\varphi=n/V$ is the concentration of monomers.  The free energy \textit{per unit
length} is similarly given by
\begin{equation}
f_K(\varphi) = - \sfrac{1}{n} \log p_{n,L}(K) .
\label{88}
\end{equation}
In figure \ref{figf01} estimates of $f_K(\varphi)$ of lattice unknots $(K=0_1)$
in $L$-cubes for $L\in\{6,8,10,12,14\}$ are plotted as a function of the 
concentration $\varphi$. In references \cite{JvR19,JvR20} these free energies, and the 
lattice osmotic pressure were examined using a Flory-Huggins approximation
\cite{Flory42,Flory53,H42}.  In these models the appropriate Flory-Huggins
approximation is a mean-field model and a function of the concentration
$\varphi$ given by
\begin{eqnarray}
f_{fh}(\varphi) = (1{-}\varphi)\log (1{-}\varphi) + (\varphi/n)\log \varphi - \chi\,\varphi^2 + A \varphi .
\end{eqnarray}
Here, $\chi$ is the Flory Interaction Parameter \cite{Flory42,H42,Flory53},
$n$ is the degree of polymerization (length of the polymer)
and the term $A\varphi$ accounts for a (bulk) entropic 
contribution to the free energy of mixing due to conformational degrees of 
freedom of the lattice polymer.  Taking first $n\to\infty$, and then
$\varphi\to 0^+$, show that $(f_{hf}(\varphi)) / \varphi \to A{-}1$.  Thus 
$A=1-\log\mu$, where $\log \mu$ is the connective constant of the
lattice (see reference \cite{JvR19} for more details).  Figure \ref{G1} is 
a schematic diagram of $f_{fh} (\varphi)$.  

The free energy curves in figure \ref{figf01} are convex functions of $\varphi$
passing through minima at concentrations $\varphi^*$.  In figure \ref{figf01} these 
minima are marked by bullets, and they decrease with increasing $L$ (size 
of the containing $L$-cube).

In the Flory-Huggins approximation a critical concentration $\varphi_c$ 
(see figure \ref{G1}) is determined by the thermodynamic stability criteria
$\frac{\partial^2}{\partial\varphi^2} f_{fh}(\varphi)=0$ and
$\frac{\partial^3}{\partial\varphi^3} f_{fh}(\varphi)=0$, taken at fixed $n$
(and temperature) (see reference \cite{ALS24} for a good review). These give 
the approximate location of a critical concentration $\varphi_c = 1/(1{+}\sqrt{n_c})$
where $n_c$ is the length of the lattice polymer at the critical concentration.   
Since $\varphi_c=n_c/(L{+}1)^3$ this expression is a self-consistent equation for 
the critical concentration $\varphi_c$.  For large $L$ this gives
$\varphi_c \approx 1/(1+\sqrt{\varphi_c\,(L{+}1)^3})$, and assuming that 
$1\ll \sqrt{(L{+}1)^3}\ll (L{+}1)^3$, one approximates $\varphi_c \approx 
1/\sqrt{\varphi_c (L{+}1)^3}$ or $\varphi_c^{3/2} \approx 1/\sqrt{(L{+}1)^3}$.  
Squaring both sides and taking the cube root gives the (Flory-Huggins) 
critical concentration as a function of $L$:
\begin{equation}
\varphi_c \approx 1/(L{+}1),
\label{eqn9a}
\end{equation} 
where $(L{+}1)^3$ is the volume (total number of lattice sites) in the containing
$L$-cube.  For example, if $L=6$, then $\varphi_c \approx 1/7 = 0.1428\ldots$.
While not equal to, this approximates the minimum in the free energy
curve for $L=6$ in figure \ref{figf01} well (for $L=6$ the minimum is located
at $\varphi^*\approx 0.169$).  Similar observations are valid for the other
values of $L$, for example, if $L=12$ then $\varphi_c \approx 1/13 = 0.0769\ldots$
while $\varphi^*\approx 0.0792$ and if $L=14$ then $\varphi_c\approx 1/15
= 0.0667\ldots$ while $\varphi^* \approx 0.067$.  Notice that $\varphi_c \approx
1/(L{+}1)$ also corresponds to $n_c \approx (L{+}1)^2$.  Following
the analysis in reference \cite{ALS24} the critical concentration
divides the free energy curves into two regimes -- for concentrations
$\varphi<\varphi_c$ the model is in a \textit{solvent-rich} regime where the lattice
polymer has self-avoiding walk exponents, and for $\varphi>\varphi_c$ the
model is in a \textit{polymer-rich} regime where the lattice polymer is
in a collapsed state, compressed into this phase by being confined in the
containing cube.  See for example figure 6 in reference \cite{ALS24}.


In this paper our focus is on the entanglements of the polymer as a function
of the concentration $\varphi$, namely as it increases from the solvent-rich
regime to the polymer-rich regime when $\varphi$ passes through its critical
value at $\varphi_c$.  Simulations in reference \cite{JOTW22} suggest that
the degree of entanglement complexity of two lattice knots increases
as the model is taken through the $\theta$-point into the collapse phase.
This result is consistent with numerous earlier studies 
\cite{AVTSR02,DERZ18,ERZ21,TJvROSW94}.
Consistent with these studies, one expects an increase in the self-entanglement
of the lattice polymer with $\varphi$, and we examine this in our model by
estimating knot probability ratios.

\subsection{Knotting probabilities in confined lattice knots} 

The probability that a randomly placed lattice polygon of length $n$ in 
an $L$-cube is a knot of type $K$ is defined by
\begin{equation}
{\mathbb P}_{n,L} (K) = \frac{p_{n,L}(K)}{p_{n,L}} .
\end{equation}
In the event that $L\gg n$, $p_{n,L} \simeq L^3\,p_n$ and 
$p_{n,L} (K) \simeq L^3\,p_n(K)$.  This shows that
\begin{equation}
\lim_{L\to\infty} {\mathbb P}_{n,L} (K) = \frac{p_n(K)}{p_n} .
\end{equation}
Since $\mu_K < \mu$ \cite{SW88} it follows by equations
\Ref{eqn3a} and \Ref{eqn3} that taking $n\to\infty$,
\begin{equation}
\lim_{n\to\infty} \lim_{L\to\infty} 
{\mathbb P}_{n,L} (K) = 0 .
\end{equation}
That is, in the limit of very large cubes the probability that a lattice 
knot has a specified knot type $K$ approaches zero.  This is valid
for any knot type $K$, including the unknot $0_1$.  Interchanging the
order of the limits above, putting $N=2\lfl L^3/2 \rfl$ one arrives at
\begin{equation}
\lim_{L\to\infty} \lim_{n\to N} 
{\mathbb P}_{n,L} (K) = \pi_K,
\end{equation}
which may be interpreted as the limiting probability that a polygon 
of maximum length $N$ in a lattice cube has knot type $K$.  In the case that
$L$ is taken to infinity through even numbers, this is the limiting knotting
probability of Hamiltonian cycles of a cube as the size of the cube expands
to infinity.

In general, for finite and fixed $L$, ${\mathbb P}_{n,L}(K)$ is a function
of $L$ and of $\varphi =n/V$, (the \textit{concentration} 
or \textit{density of monomers} in a cube of sidelength $L$).  We shall
see that numerical estimates of ${\mathbb P}_{n,L}(K)$ are not directly
accessible, since algorithms sampling lattice knots in confined geometries
are either not efficient as the concentration of occupied lattice sites becomes
high (one may, for example, use the pivot algorithm for lattice polygons
\cite{JvRW90P,JvR09MC}, but this algorithm has a very low success rate of
proposed elementary moves when the concentration of occupied sites is high).
Other algorithms are also inefficient or are only usefull to sample polygons of
fixed knot type in confining spaces (see references \cite{JvR19,JvR20}), and 
cannot directly be used to determine knotting probabilities.

Since the GARM and GAS algorithms can be used to determine 
approximate counts of states when sampling lattice clusters, it is 
possible to estimate ratios of knotting probabilities.  That is, estimates of
\begin{equation}
\rho_{n,L}(K_1/K_2) = \frac{p_{n,L}(K_1)}{p_{n,L}(K_2)}
= \frac{{\mathbb P}_{n,L}(K_1)}{{\mathbb P}_{n,L}(K_2)} 
\label{12}
\end{equation}
can be obtained by taking ratios of the estimates of $p_{n,L}(K)$ for
various knot types $K$ using the GARM or GAS algorithms.  In the
case that the lattice knots fill the $L$-cube, define
\begin{equation}
\scalebox{0.95}{$\displaystyle
R_L(K_1/K_2)=
\lim_{n\nearrow V} \rho_{n,L}(K_1/K_2) 
= \lim_{n\nearrow V} \frac{{\mathbb P}_{n,L}(K_1)}{{\mathbb P}_{n,L}(K_2)} .$}
\label{e16}
\end{equation}
This is the (limiting) ratio of probabilities of knot types $K_1$ and $K_2$ as the
polygon approaches a Hamiltonian cycle inside the cube.  The ratio
$\rho_{n,L}(K_1/K_2)$ is generally a function of the concentration of
monomers $\varphi$, and its size indicates the relative incidence of knot type
$K_1$ (rather than knot type $K_2$) if states are sampled uniformly.
That is, if $\rho_{n,L}(K_1/K_2)<1$ then one expects the knot type
$K_2$ to be realised in lattice knots inside the confining cube more
abundantly than knot type $K_1$.  This quantity will be examined for
various knot types in the next sections of this paper.   In particular,
the case that $K_2=0_1$ gives the ratio of a knot type $K_1$ compared
to the unknot.  If $R_L(K_1/0_1) < 1$, then the knot type $K_1$ is
suppressed compared to the unknot, and this is what the data will
show when $L$ is finite.

\section{Sampling lattice knots confined to cubes in the cubic lattice}

The GAS \cite{JvRR09} and GARM \cite{RJvR08} algorithms were 
implemented using BFACF elementary moves \cite{BF81,ACF83} to 
sample lattice knots inside an $L$-cube of fixed size.  Normalising the 
results using the data in table \ref{table1}, approximate counts of the number of
distinct lattice knots of length $n$ (with $n_K \leq n \leq L^3$) can be obtained.
The implementation of these algorithms was along $4$ parallel threads
(and in some cases, $6$ parallel threads) using a parallel implementation 
similar to that of parallel PERM \cite{CJvR20}.  Our results were obtained using
the GAS implementation and additional short simulations were done 
using the GARM algorithm to verify our results.

Each GAS simulation was done in $B$ blocks, each block consisting of 
$T$ parallel threads, and each thread with $S$ iterations. For example, in a
$12$-cube with $2197$ sites, lattice knots of length $n\leq 2196$ were sampled
using $B=120$ blocks, each block having $T=4$ parallel threads and 
each call to a thread having  $S=20{,}000{,}000$ BFACF iterations (for a total
of $BST=9{,}600{,}000{,}000$ BFACF elementary moves collecting data on lattice 
knots for of lengths between $n_K=4$ and maximum length $2196$ 
in the case that $K=0_1$).  The number of blocks increased from $60$ for $L=6$ 
to $150$ for $L=14$, and the lengths of threads from $20{,}000{,}000$ for $L=6$ to 
$30{,}000{,}000$ for $L=14$.

Since the data collected in each block becomes independent from other blocks, 
rough estimates of the error bar on the results can be obtained.  While the 
error is relatively small, it increases in size with the length of the lattice knot but shows that the order of magnitude of the approximate counts is
accurate.  For example, if $L=12$ and the lattice unknot is sampled, then
for $n=10$ on gets $p_{10,12}(0_1) = (8.55\pm0.28)\times 10^5$
and $p_{1000,12}(0_1) = (3.04\pm1.54)\times 10^{563}$.  While the
error bar in the last estimate appears large, it is the case that 
$\log p_{1000,12}(0_1)=1297.5 \pm 1.2$, with an acceptable error bar.

\begin{figure}[h!]
\includegraphics[width=0.5\textwidth,height=0.33\textheight]{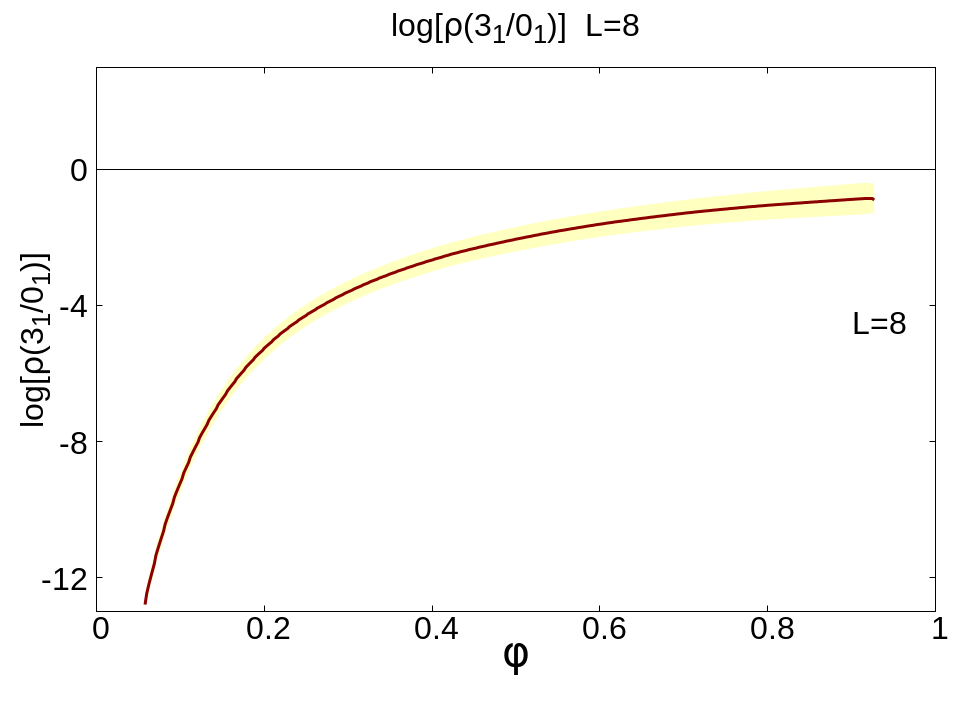}
\captionsetup{justification=raggedright,singlelinecheck=false}
\caption{The knot probability ratio $\rho_{n,L}(3_1^+/0_1)$ of trefoils to 
unknots for $L=8$ (see equation \Ref{12}) plotted against the concentration
$\varphi$.  Notice the logarithmic scale and that trefoils are rare compared 
to unknots in this model of compressed lattice knots.  With increasing 
concentration the ratio increases monotonically, but remains small.
The confidence interval is shown by the shaded area around the curve
and was determined by combining the standard deviations calculated for
the raw data and then combined via the ratio in equation \Ref{12}.}
\label{figu3}
\end{figure}

The algorithms were implemented and data were collected through several
months, resulting in data being available for $L$-cubes of sizes 
$L\in\{6,8,10,12\}$ and also $L=14$ for the unknot and trefoil knot types.
Lattice knots sampled in $12$-cubes had lengths up to $13^3-1=2196$,
but convergence was generally good only for concentrations $\varphi \lessapprox 0.93$.
This was similarly the case for sampling in $14$-cubes (where lattice knots
of lenght up to $15^3-1=3374$ were sampled, but convergence was
achieved when $\varphi \lessapprox 0.90$).  Convergence for smaller values of
$L$ was generally better, resulting in shorter simulations, although sampling
for high values of $\varphi$ remained challenging.  Generally, for 
concentrations $\varphi \gtrapprox 0.93$ the ratios $\rho_{n,L}(K_1/K_2)$ 
were noisy and no conclusions could be drawn.

\subsection{The relative incidence of unknots and trefoils}

\begin{figure}[h!]
\includegraphics[width=0.5\textwidth,height=0.33\textheight]{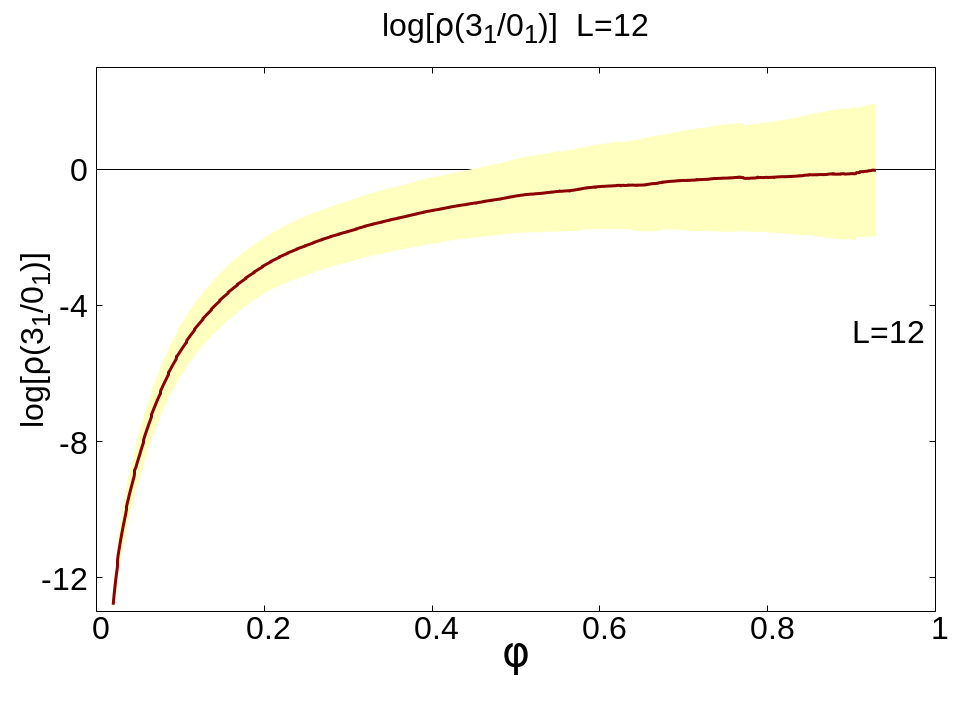}
\captionsetup{justification=raggedright,singlelinecheck=false}
\caption{The knot probability ratio $\rho_{n,L}(3_1^+/0_1)$ of trefoils to 
unknots for $L=12$ (see equation \Ref{12}) plotted against the concentration
$\varphi$.  The scaling along the axes is the same as in figure \ref{figu3},
and with increasing concentration, the ratio increases monotonically, although
it is still small in this larger confining cube at low concentrations but increases to
$1$ as the concentration increases beyond $0.8$. The confidence 
interval is again shown by the shaded area around the curve.}
\label{figu4}
\end{figure}

Figure \ref{figu3} is a plot of $\rho_{n,L}(3_1^+/0_1)$ as a function of $\varphi$ 
for $L=8$.  The knotting probability of trefoils ($3_1^+$), relative to the unknot 
($0_1$), increases monotonically with the concentration $\varphi$ and 
eventually levels off for sufficiently large values of $\varphi$.  The shaded 
band around the curve is a standard deviation confidence interval.  At 
$\varphi\approx 0.5$ the relative knotting probability is roughly $0.10$ 
indicating that the number of trefoil lattice knots is approximately only 
$10$\% of the corresponding unknots ($0_1$).  In other words, compared to unknotted
lattice knots, trefoils are relatively rare.  Similar data are shown for $L=12$ in 
figure \ref{figu4}.  In this case the relative incidence of trefoils compared to unknots 
is larger compared compared to inknots, increasing from low values (less than 
$0.10$ for $\varphi<0.3$) to approach $1.0$ when $\varphi>0.8$  The 
uncertainty envelope in this figure is also wider giving a large uncertainty envelope
when $\varphi$ exceeds approximately $0.5$.  In both figures \ref{figu3}
and \ref{figu4} the curves level off as the concentration approaches $1$, 
indicating that the number of knotted polygons (of type trefoil) comes closer
to the number of unknotted polygons as the Hamiltonian limit is approached.
However, these data also show that the unknot is a more popular knot type
compared to the trefoil over the entire range of concentration, in particular
for small values of $L$.  

\begin{figure}[h!]
\includegraphics[width=0.5\textwidth,height=0.33\textheight]{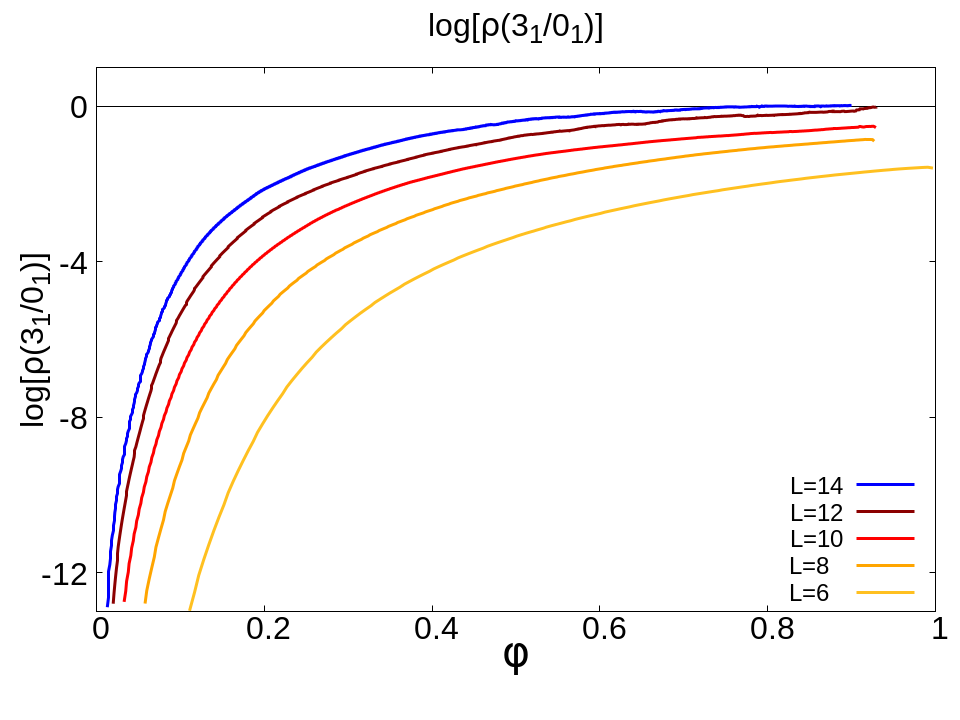}
\captionsetup{justification=raggedright,singlelinecheck=false}
\caption{The knot probability ratio $\rho_{n,L}(3_1^+/0_1)$ of trefoils to 
unknots for $L\in\{6,8,10,12,14\}$ plotted against the concentration
$\varphi$.  The scaling along the axes is the same as in figure \ref{figu3}.  With
increasing $L$ the ratio increases but also suggests that the ratio may be
approaching a limiting curve less than $1$ but approaching $1$ as the
concentration approaches $1$.}
\label{figu5}
\end{figure}

In figure \ref{figu5} the $\varphi$ dependence of the trefoil-unknot ratio 
$\rho_{n,L}(3_1^+/0_1)$ is reported for several confining $L$-cubes with 
$L\in\{6,8,10,12,14\}$.  For fixed concentration $\varphi$ the ratios increase 
systematically with $L$,  and there is also a levelling off at larger values of $L$.
The curves appear to approach a limiting curve (especially if the increase in volume
of the $L$-cube proportional to $O(L^3)$ is taken into account).  However, 
even if the data are suggestive,  the numerical simulations at larger values 
of $L\geq 12$ pose significant challenges.  If these curves approach a limiting 
value less than $1$ (that is, if  $\limsup_{L\to\infty} \rho_{n,L}(3_1^+/0_1) < 1$ 
when the limsup is taken at constant concentration), then trefoils remain 
suppressed compared to unknots, even in the continuum limit. However, the
data appear to show that the curves accumulate, with increasing $L$, on
the horizontal axis, so that $\rho_{n,L}(3_1^+/0_1) \to 1$ as $L\to\infty$ for
Hamiltonian polygons.

Note that the probability ratios in figure \ref{figu5} increase with the 
concentration $\varphi$, and sharply so as $\varphi$ transitions from below 
$\varphi_c\approx 1/(L{+}1)$  (the solvent-rich regime) to above $\varphi_c$ 
in the polymer-rich or collapsed regime.  This is consistent with other studies 
on lattice knot probabilities. Lattice knots are generally smaller the more 
complex the knot type is \cite{JvRW91a}.  Therefore, more complex knot types
are more likely to occur when a polygon is compressed \cite{GJvRR12,AVTSR02}, 
or when the lattice polygon is in a more concentrated (collapsed) 
regime \cite{TJvROSW94,DEMRZ14,DERZ18,ERZ21}. 

Additionally, the curves in figure \ref{figu5} flatten out as the concentration 
increases above $0.7$.  This shows that the incidence of knots of type $3_1^+$ 
becomes stable when compared to unknots, suggesting that there is a stable 
ratio of lattice knots of unknot and trefoil types for fixed $\varphi$ that increases
approaching the Hamiltonian state.  In this case one would be interested in 
the limiting ratio
\begin{equation}
R(3_1^+/0_1) = \lim_{L\to\infty} \lim_{\varphi\to 1^-} \rho_{n,L}(3_1^+/0_1) \leq 1.
\end{equation}
Interesting questions to be addressed are whether $R(3_1^+/0_1)<1$ and if 
its value is universal (that is, independent of the model, and so an intrinsic 
property of the knot types for dense curves).

The curves in figure \ref{figu5} appear to start converging with increasing 
$\varphi$, but the accessible data point is only suggestive.  Only simulations 
at larger values of $L$, and successful collection of data for $\varphi$ close 
to $1$, would settle these questions, but are also outside the realm of what 
is possible using our current algorithms and computer hardware.

\begin{figure}[h!]
\includegraphics[width=0.5\textwidth,height=0.33\textheight]{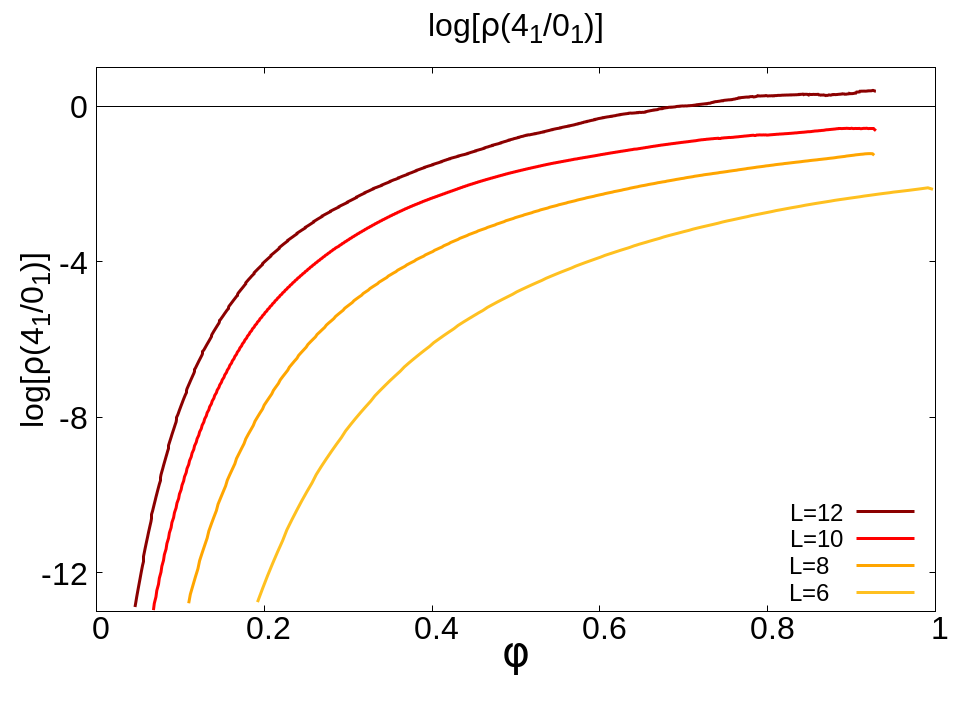}
\captionsetup{justification=raggedright,singlelinecheck=false}
\caption{The relative knot probability ratio $\rho_{n,L}(4_1/0_1)$ of figure eight
knots ($4_1$) to unknots for $L\in\{6,8,10,12\}$ plotted against the concentration
$\varphi$.  The scaling along the axes is the same as in figure \ref{figu3}.}
\label{figu6}
\end{figure}

\subsection{The relative incidence of other prime knots}

The knot probability ratios $\rho_{n,L}(K/0_1)$ of other prime knots of type 
$K$ with respect to the unknots were also estimated by collecting data on 
the figure eight knot ($4_1$), and five and six crossing prime knots in 
the standard knot tables (namely knot types $5_1$ and $5_2$, and 
the six crossing knots $6_1$, $6_2$ and $6_3$ and the compound
six crossing knots $3_1^+\#3_1^+$ and $3_1^-\# 3_1^+$).  In 
these cases lattice knots were sampled in confining cubes of side-lengths 
$L\in\{6,8,10,12\}$, and again scaled using the data in table \ref{table1}.

\begin{figure}[h!]
\includegraphics[width=0.5\textwidth,height=0.33\textheight]{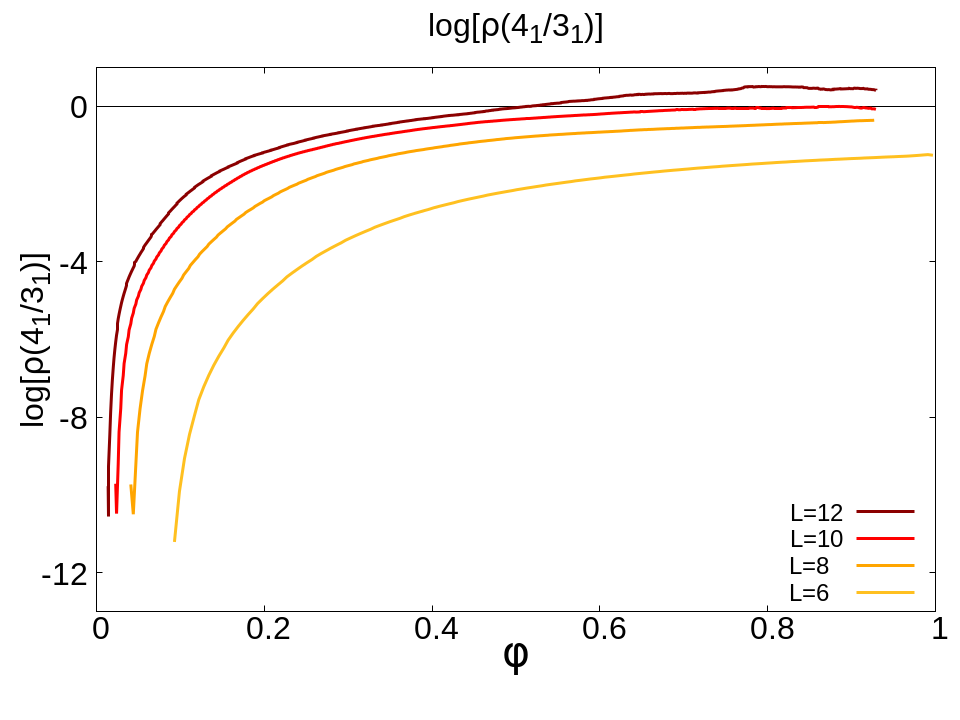}
\captionsetup{justification=raggedright,singlelinecheck=false}
\caption{The relative knot probability ratio $\rho_{n,L}(4_1/3_1^+)$ of 
figure eight knots ($4_1$) to the trefoil knot ($3_1^+$) for $L\in\{6,8,10,12\}$ 
plotted against the concentration $\varphi$.  The scaling along the axes is 
the same as in figure \ref{figu3}.}
\label{figu7}
\end{figure}

In figure \ref{figu6} the relative knot probability ratios $\rho_{n,L}(4_1/0_1)$ of 
the figure eight ($K=4_1$) are shown.  Comparing the curves in figure \ref{figu6} to 
those reported in figure \ref{figu5} for knot type $3_1^+$  shows that the ratios for $4_1$
are lower than the corresponding ratios for $3_1^+$,  with the exception of the 
curve for $L=12$ at high concentration (examination of our data shows that this
curve is more noisy at large concentration).  This indicates that, generally,  $4_1$ lattice 
knots are less likely to occur than the corresponding lattice knots of knot type
$3_1^+$.  Note that the curves in figure \ref{figu6} do have the same general 
behaviour as seen for $3_1^+$ in figure \ref{figu5}, but they do increase more 
sharply as the concentration $\varphi$ increases from the solvent-rich 
regime $\varphi<\varphi_c$ to the polymer-rich regime $\varphi>\varphi_c$.

The relation between $3_1^+$ and $4_1$ may be better explored by plotting the 
$\varphi$ dependence of the knot probability ratio $\rho_{n,L}(4_1/3_1^+)$, see 
figure \ref{figu7}.  In these cases the data are consistent with $\rho_{n,L}(4_1/3_1^+)<1$
generally (except for $L=12$ at high concentration where the curve appears more
noisy), indicating that the probability of $4_1$ is generally less than that of 
$3_1^+$.  As $L$ increases, the ratio $\rho_{n,L}(4_1/3_1^+)$ appears to approaches 
$1$ (the confidence interval, as suggested in figures \ref{figu3} and \ref{figu4} 
widens with increasing concentration, and the horizontal axis is well within 
it at high concentration).  This suggests that the incidence of knot types 
$4_1$ and $3_1^+$ may become similar for large $L$, at least as the concentration 
approaches $1$.  However, for low concentrations it is not evident that 
$\rho_{n,L}(4_1/3_1^+)$ is approaching $1$ as $L$ increases, but this cannot be ruled out.

\begin{figure}[h!]
\includegraphics[width=0.5\textwidth,height=0.33\textheight]{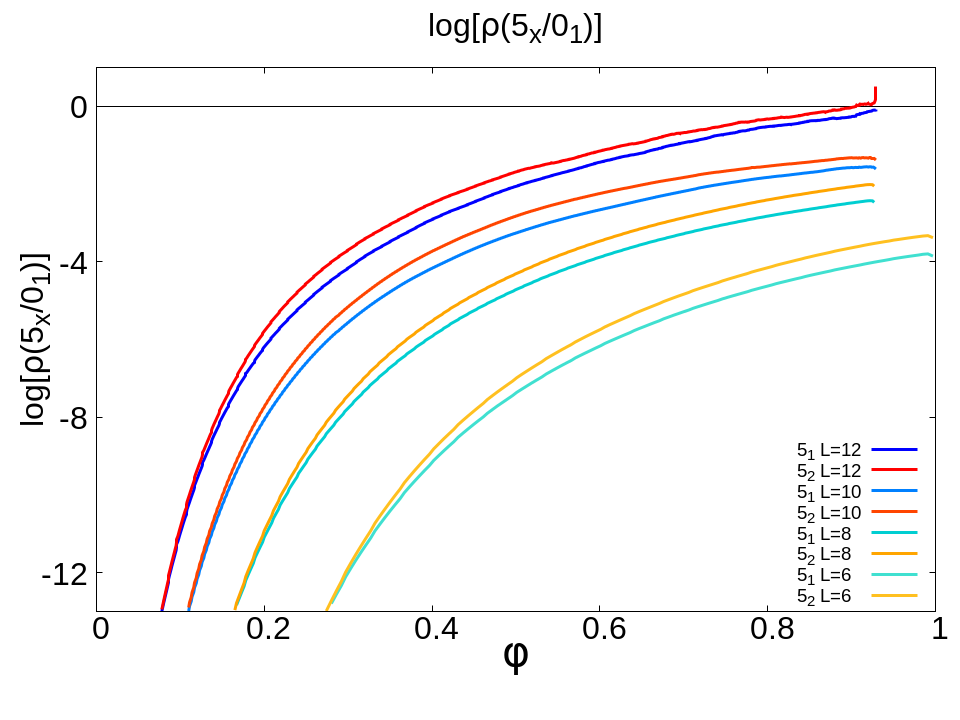}
\captionsetup{justification=raggedright,singlelinecheck=false}
\caption{The relative knot probability ratio $\rho_{n,L}(5_x/0_1)$ of figure eight
knots ($5_x$) to unknots for $L\in\{6,8,10,12\}$ plotted against the concentration
$\varphi$.  The scaling along the axes is the same as in figure \ref{figu3}.}
\label{figu8}
\end{figure}

\begin{figure}[h!]
\includegraphics[width=0.5\textwidth,height=0.33\textheight]{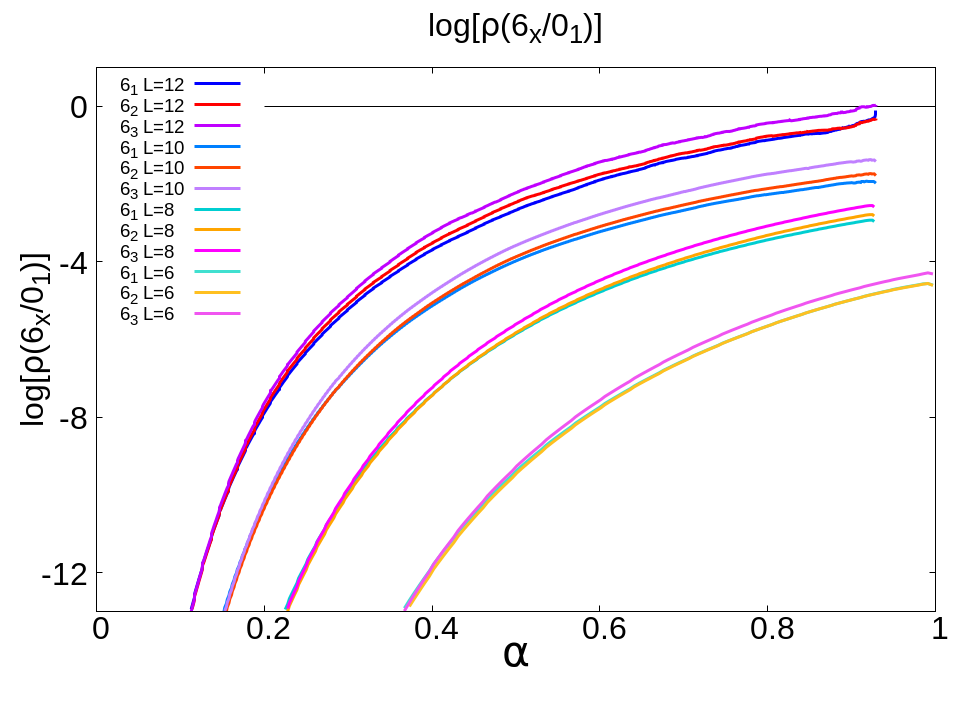}
\captionsetup{justification=raggedright,singlelinecheck=false}
\caption{The relative knot probability ratio $\rho_{n,L}(6_x/0_1)$ of figure eight
knots ($6_x$) to unknots for $L\in\{6,8,10,12\}$ plotted against the concentration
$\varphi$.  The scaling along the axes is the same as in figure \ref{figu3}.}
\label{figu9}
\end{figure}

\begin{figure}[t!]
\includegraphics[width=0.5\textwidth,height=0.33\textheight]{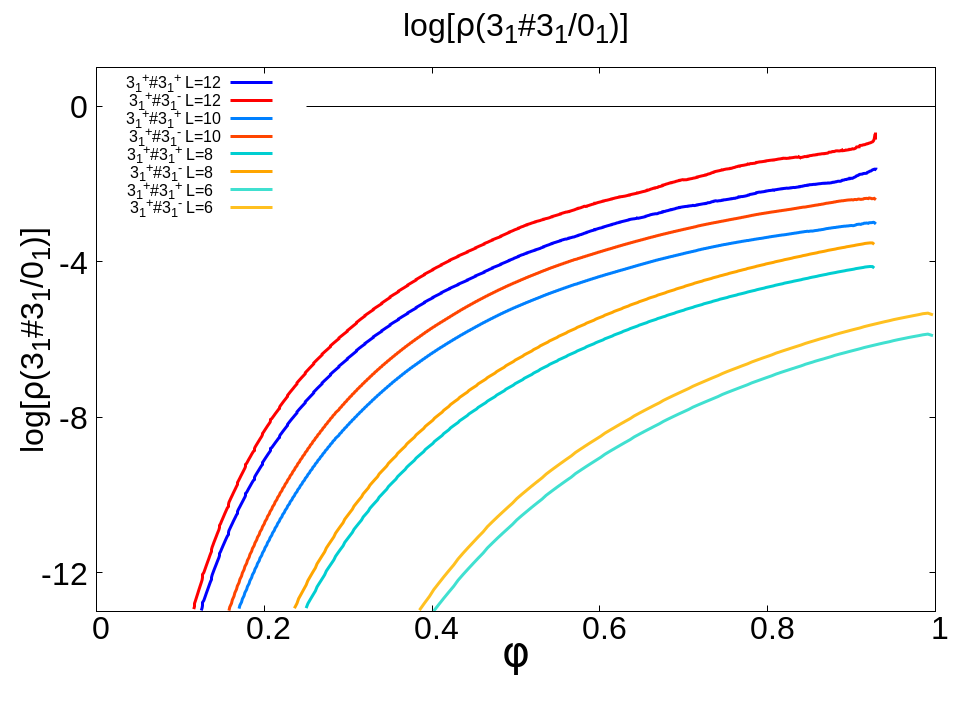}
\captionsetup{justification=raggedright,singlelinecheck=false}
\caption{The relative knot probability ratio $\rho_{n,L}(3_1\#3_1/0_1)$ 
for two compounded trefoil knots of types $3_1^+\#3_1^+$ and
$3_1^+\#3_1^-$ to unknots for $L\in\{6,8,10,12\}$ plotted against 
the concentration $\varphi$.  The scaling along the axes is the same as in figure \ref{figu3}.}
\label{figu10}
\end{figure}

In figure \ref{figu8} the relative probabilities of the five crossing knots $5_1$ 
and $5_2$ with respect to the unknot $0_1$ are plotted against concentration 
for different values of $L$.  One notices that the incidence of five-crossings 
knots is suppressed compared to the trefoil or figure eight knots. Moreover, knot 
type $5_2$ is more likely to occur, compared to  $5_1$.  Similar plots are reported 
in figure \ref{figu9} for the six crossing knot types $6_1$, $6_2$ and $6_3$.  
Of these three knot types, $6_3$ is the most likely, followed by $6_2$ and 
$6_1$.  Finally, the relative probabilities of the compound knots $3_1^+\#3_1^-$ 
and $3_1^+\#3_1^+$ are shown in figure \ref{figu10}, and $3_1^+\# 3_1^-$ 
has a higher probability of occurring.

\begin{table}[t!]
\begin{threeparttable}
\caption{Estimates of $\log p_{n,L}(K)$ in an $L$-cube with $L=12$ in descending order}
\setlength{\extrarowheight}{3pt}
\setlength{\tabcolsep}{6pt}            
\begin{tabular}{lrrr}
\hline
Knot type & $n=500$ & $n=1000$ & $n=2000$ \\
\hline
$0_1$ & $699.69(45)$ & $1297.48(52)$ & $1816.77(93)$ \\
$3_1^+$ & $697.22(41)$ & $1296.51(50)$ & $1816.71(99)$ \\
$4_1$ & $696.23(43)$ & $1296.36(52)$ & $1817.98(160)$ \\
$5_2^+$ & $694.64(45)$ & $1295.97(54)$ & $1816.83(76)$ \\
$5_1^+$ & $694.19(37)$ & $1295.06(46)$ & $1816.59(81)$ \\
$6_3$ & $693.02(49)$ & $1294.81(62)$ & $1816.75(99)$ \\
$6_2^+$ & $692.86(47)$ & $1294.56(52)$ & $1816.36(81)$ \\
$6_1^+$ & $692.74(57)$ & $1294.38(61)$ & $1816.13(79)$ \\
$3_1^{+}\#3_1^{-}$ & $692.24(45)$ & $1291.56(55)$ & $1815.77(75)$ \\
$3_1^{+}\#3_1^{+}$ & $691.49(65)$ & $1293.15(72)$ & $1815.02(91)$ \\
\hline
\end{tabular}
\label{t2}
\end{threeparttable}
\end{table}

The relative knot probability of knot types for $L=12$ are ranked in
table \ref{t2} by listing $\log p_{n,L}(K)$ in decreasing size for $n=500$.
The data for $n=1000$ and $n=2000$ are also listed for comparison,
and are generally consistent with the ranking at $n=500$, except in 
a few cases.  For example, although not separated by the error
bars, the knot type $4_1$ appears to outrank the unknot and
$3_1^+$ when $n=2000$, but a large confidence interval shows that
this is likely due to uncertainties in the data.  However, the general 
trend in the table is that increasing the crossing number suppresses the
number of conformations of a lattice knot, and so reducing its
probability of being sampled if lattice knots are sampled uniformly.
This is in particular the case for small values of $L$.  Generally,
the unknot outranked all the non-trivial knot types, and by a wide margin
at low concentration.  It is not clear that this observation will
persist with increasing $L$ and concentration, as shown in table
\ref{t2} and in the graphs of $\rho_{n,L}(K/0_1)$ in figures \ref{figu5}
to \ref{figu8}.

\section{Conclusions}

Our numerical results show that the unknot is the most popular knot type
when lattice knots are compressed in an $L$-cube, for values of $L$ up to $L=14$.
This is in particular the case for low to moderately high concentrations.
In figure \ref{figu5} the curves of knot probabilities of the trefoil relative to 
the unknot suggest that the trefoil will remain rare compared to unknots
for small values of $L$. This is less clear with increasing $L$; while the rate 
of change in the curves of $\rho_{n,L}(3_1^+/0_1)$ decreases with increasing 
$L$, these curves may reach a limiting curve at zero.  Additionally, as the concentration
$\varphi$ increases the relative knotting probability, while still increasing, 
flattens out to a limiting value less than zero as the Hamiltonian state 
is reached for smaller values of $L$.  These observations also apply to the other 
knot types examined. However, we expect that unknots will become rare as $L$ 
increases, compared to all non-trivial knot types, but the possibility that 
the unknot is dominant compared to any other fixed non-trivial knot types
appear likely for low concentration.  That is, in the limit that $L$ is large, 
and polygons are sampled uniformly from a confining cube, one expects 
non-trivial knot types to be far more likely than unknots.  However, unknots may 
still be the single most likely knot type, consistent with our numerical results.

Additionally, the likelihood of knotted conformations in an $L$-cube increases 
with $\varphi$ as the model passes through the critical concentration 
$\varphi_c$ separating the solvent-rich and polymer-rich regimes.  Thus, an 
increase in knotting (and thus entanglement) is expected as the concentration
increases towards the limit of Hamiltonian cycles of the cube.  In other words, 
in the dense phase entanglements, as measured by the knotting of the 
lattice polygons, is strongly enhanced, 

\begin{table}[h]
\begin{threeparttable}
\caption{Limiting probability ratio $R_L(K_1/K_2)$ (equation \Ref{e16})}
\setlength{\extrarowheight}{3pt}
\setlength{\tabcolsep}{1pt}            
\begin{tabular}{lrrrrr}
\hline
$L$ & $3_1^+/0_1$ & $4_1/0_1$ & $4_1/3_1^+$ & $5_1^+/0_1$ & $5_2^+/0_1$ \\
\hline
$6$ & $-1.538(60)$ & $-2.030(9)$ & $-1.507(54)$ & $-3.689(13)$ & $-3.197(14)$ \\ 
$8$ & $-0.832(23)$ & $-1.059(22)$ & $-0.404(21)$ & $-2.185(29)$ & $-1.765(39)$ \\
$10$ & $-0.572(36)$ & $-0.377(36)$ & $-0.137(31)$ & $-1.391(42)$ & $-1.049(45)$ \\
$12$ & $-0.391(34)$ & $0.291(42)$ & $-0.051(29)$ & $-0.363(48)$ & $-0.070(51)$ \\
$14$ & $-0.301(24)$ \\
\hline
\end{tabular}
\label{t3}
\end{threeparttable}
\end{table}
\def\w{\textcolor{white}{$-$}}
\begin{table}[h]
\begin{threeparttable}
\caption{Limiting probability ratio $R_L(K_1/0_1)$ (equation \Ref{e16}) for $6$ crossing knots}
\setlength{\extrarowheight}{3pt}
\setlength{\tabcolsep}{1pt}            
\begin{tabular}{lrrrrr}
\hline
$L$ & $6_1^+/0_1$ & $6_2^+/0_1$ & $6_3/0_1$ & $3_1^+\#3_1^+/0_1$ & $3_1^+\#3_1^-/0_1$ \\
\hline
$6$ & $-4.424(27)$ & $-4.416(19)$ & $-4.152(23)$ & $-5.736(18)$ & $-5.200(22)$ \\ 
$8$ & $-2.602(47)$ & $-2.460(47)$ & $-2.216(52)$ & $-3.796(50)$ & $-3.190(46)$ \\
$10$ & $-1.655(54)$ & $-1.481(55)$ & $-1.113(57)$ & $-2.727(63)$ & $-2.100(54)$  \\
$12$ & $-0.431(59)$ & $-0.330(58)$ & $0.009(63)$ & $-1.670(66)$ & $-0.983(59)$  \\
\hline
\end{tabular}
\label{t4}
\end{threeparttable}
\end{table}

In table \ref{t3} we plot the logarithms of the limiting probability ratios 
$R_L(K_1/K_2)$ as defined in equation \Ref{e16}.  These ratios correspond to 
the extrapolation of $\rho_L(K_1/K_2)$ to $\varphi=1$ (in other words, to a space 
filling (Hamiltonian) curve in the confining cube).  For example, if the curve 
in figure \ref{figu3} is extrapolated then $\log R_8(3_1^+/0_1) = -0.832 \pm 0.023$.
The extrapolation was done assuming the model
\begin{eqnarray}
\log \rho_L(K_1/K_2) &=& \log R_L(K_1/K_2) \\
& & + b\left(1/\varphi - 1\right) + c\left(1/\varphi^2 - 1\right) \nonumber
\end{eqnarray}
and performing a weighted least squares fit and discarding data at small values
of $\varphi$ (typically, we fitted against data for $\varphi>0.1$).

In table \ref{t3} there is a systematic decrease in the size of $\log R_L(K_1/K_2)$ 
down the columns, showing that non-trivial knot types generally become more 
competitive with the unknot as $L$ increases.  However, the dominance of
the unknot remains substantial for smaller values of $L$, even as the limiting ratio
$\log R_L(4_1/3_1^+)$ approaches zero with increasing $L$ (this indicates that
the probability of knot type $4_1$ may become comparable to that of
the trefoil knot type $3_1^+$ as the size of the confining cube increases). 

The variation along rows in table \ref{t3} are more uneven, indicating
that some knot types are more likely to occur than others.  For example,
comparing the knot types $5_1$ and $5_2$ shows that $5_2$ is more
likely than $5_1$ in the limit that $\varphi\nearrow 1$.  Similar observations
are valid for the knot types $3_1^+$ and $4_1$.  In table \ref{t4} all the
six crossing prime and compound knot types are listed.  An examination
of the results show that generally, except for $L=6$, amongst the
six crossings knots the knot type $6_3$ the most likely. The six
crossing compound knot types $3_1^\pm\#3_1^\pm$ are suppressed
compared to the prime six crossing knots, and with the knot type
$3_1^+\#3_1^-$ more likely than $3_1^+\#3_1^+$ (or $3_1^-\#3_1^-$).

\section*{Acknowledgements} EJJvR acknowledges financial support 
from NSERC (Canada) in the form of Discovery Grant RGPIN-2019-06303. 
MCT acknowledges membership of GNAMPA 
(Gruppo Nazionale per l’ Analisi Matematica, la Probabilità  e le loro Applicazioni) 
of INdAM (Istituto Nazionale di Alta Matematica), Italy.

\bibliographystyle{unsrt}
\bibliography{confineknots}

\end{document}

%% file: macros
\def\2;{\;\;}

\def\IntZ{{\mathbb Z}}

\def\w#1{\widetilde{#1}}
\def\Ref#1{(\ref{#1})}

\def\sfrac#1#2{\hbox{\nor $\frac{#1}{#2}$}}

\def\lfl{\!\left\lfloor} \def\rfl{\right\rfloor\!}

\def\nor{\normalsize}

